\documentclass[prd, aps, floatfix, 10pt, twocolumn, superscriptaddress, preprintnumbers,nofootinbib]{revtex4-1}

\usepackage[utf8]{inputenc}
\usepackage{xcolor}

\definecolor{lcolor}{rgb}{0.5,0,0}
\definecolor{citcolor}{rgb}{0,0.3,0.0}

\usepackage[breaklinks,colorlinks,urlcolor=blue,citecolor=citcolor,linkcolor=lcolor]{hyperref}

\usepackage{amsmath}
\usepackage{physics}
\usepackage{siunitx}
\usepackage{slashed}

\usepackage{graphicx}

\usepackage{mathtools}

\usepackage{overpic}

\usepackage{subfigure}

\usepackage{tikz}
\usepackage{tikz-3dplot}
\usetikzlibrary {arrows.meta}
\usetikzlibrary{shapes.misc}

\tikzset{cross/.style={cross out, draw=black, minimum size=2*(#1-\pgflinewidth), inner sep=0pt, outer sep=0pt},cross/.default={1pt}}
\usetikzlibrary{decorations.pathmorphing}
\usetikzlibrary{snakes,decorations.pathmorphing}

\newcommand{\ef}{\epsilon_f}

\newcommand{\aem}{\alpha_\text{em}}

\newcommand{\rt}{\mathbf{r}}
\newcommand{\kt}{\mathbf{k}}

\newcommand{\pt}{\mathbf{p}}
\newcommand{\qt}{\mathbf{q}}

\newcommand{\GeV}{{{\,}\textrm{GeV}}}

\newcommand*\diff{\mathop{}\!\mathrm{d}}

\graphicspath{ {figures/} }

\usepackage{cleveref}
\crefname{section}{Sec.}{Secs.}
\crefname{figure}{Fig.}{Figs.}
\crefname{appendix}{Appendix}{Appendices}
\crefname{equation}{Eq.}{Eqs.}
\crefname{table}{Table}{Tables}
\Crefname{section}{Section}{Sections}
\Crefname{figure}{Figure}{Figures}
\Crefname{appendix}{Appendix}{Appendices}
\Crefname{equation}{Equation}{Equations}
\Crefname{table}{Table}{Tables}

\usepackage{verbatim}

\newcommand{\addfignewdouble}[6]{
\begin{figure*}[htp!]
    \centering
    \subfigure[$\frac{1}{A}\frac{\dd{\tilde\Sigma_{#1}}}{\dd{\tau}}$]{\label{subfig:#1-#5-#6}\includegraphics[width=0.4\textwidth]{figures/plots_#1_Au_Q#5_y#6.pdf}}\qquad
    \subfigure[$\frac{1}{A}\frac{\dd{\tilde\Sigma_{#3}}}{\dd{\tau}}$]{\label{subfig:#3-#5-#6}\includegraphics[width=0.4\textwidth]{figures/plots_#3_Au_Q#5_y#6.pdf}}
    \caption{
    The OPECs and their nuclear modification factor 
    as functions of $\tau$ for $\sqrt{s}=\SI{90}{GeV}$, $y=#6$, $Q=\SI{#5}{GeV}$.
     \label{fig:Q#5_y#6}
    }
\end{figure*}
}

\newcommand{\addfignewquadruple}[9]{
\begin{figure*}[htp!]
    \centering
    \subfigure[$\frac{1}{A}\frac{\dd{\tilde\Sigma_{#1}}}{\dd{\tau}}$]{\label{subfig:#1-#8-#9}\includegraphics[width=0.4\textwidth]{figures/plots_#1_Au_Q#8_y#9.pdf}}\qquad
    \subfigure[$\frac{1}{A}\frac{\dd{\tilde\Sigma_{#3}}}{\dd{\tau}}$]{\label{subfig:#3-#8-#9}\includegraphics[width=0.4\textwidth]{figures/plots_#3_Au_Q#8_y#9.pdf}}
    \subfigure[$\frac{1}{A}\frac{\dd{\tilde\Sigma_{#5}}}{\dd{\tau}}$]{\label{subfig:#5-#8-#9}\includegraphics[width=0.4\textwidth]{figures/plots_#5_Au_Q#8_y#9.pdf}}\qquad
    \subfigure[$\frac{1}{A}\frac{\dd{\tilde\Sigma_{#7}}}{\dd{\tau}}$]{\label{subfig:#7-#8-#9}\includegraphics[width=0.4\textwidth]{figures/plots_#7_Au_Q#8_y#9.pdf}}
    \caption{
    The OPECs and their nuclear modification factor as functions of $\tau$ for $\sqrt{s}=\SI{90}{GeV}$, $y=#9$, $Q=\SI{#8}{GeV}$.
    \label{fig:Q#8_y#9}
    }
\end{figure*}
}

\begin{document}
\title{One-point energy correlator for deep inelastic scattering at small $x$}

\author{Zhong-Bo Kang}
\email{zkang@physics.ucla.edu}
\affiliation{Department of Physics and Astronomy, University of California, Los Angeles, CA 90095, USA}
\affiliation{Mani L. Bhaumik Institute for Theoretical Physics, University of California, Los Angeles, CA 90095, USA}
\affiliation{Center for Frontiers in Nuclear Science, Stony Brook University, Stony Brook, NY 11794, USA}

\author{Robert Kao}
\email{rqk@ucla.edu}
\affiliation{Department of Physics and Astronomy, University of California, Los Angeles, CA 90095, USA}
\affiliation{Mani L. Bhaumik Institute for Theoretical Physics, University of California, Los Angeles, CA 90095, USA}

\author{Meijian Li}
\email{corresponding author: meijianli@hust.edu.cn}
\affiliation{School of Physics, Huazhong University of Science and Technology, Wuhan 430074, China}
\affiliation{Instituto Galego de Fisica de Altas Enerxias (IGFAE), Universidade de Santiago de Compostela, E-15782 Galicia, Spain}

\author{Jani Penttala}
\email{janipenttala@physics.ucla.edu}
\affiliation{Department of Physics and Astronomy, University of California, Los Angeles, CA 90095, USA}
\affiliation{Mani L. Bhaumik Institute for Theoretical Physics, University of California, Los Angeles, CA 90095, USA}

\begin{abstract}

We derive the expressions for the one-point energy correlator (OPEC) in deep inelastic scattering in the high-energy (small-$x$) limit within the Color Glass Condensate (CGC) framework. The OPEC is computed as a function of the angle between the energy flow and the target proton or nucleus, enabling a systematic exploration of different momentum scales in the scattering process. Owing to the momentum sum rule, the dependence on fragmentation functions cancels, leaving the dipole amplitude as the only nonperturbative input. As a result, the OPEC provides a sensitive and promising probe of gluon saturation dynamics at small $x$. We present numerical results for representative kinematic configurations relevant to the future Electron--Ion Collider, demonstrating sizable nuclear suppression effects within the CGC framework and highlighting the consistency of this observable with predictions from gluon saturation.

\end{abstract}

\maketitle

\section{Introduction}

The energy-energy correlator (EEC), first proposed in the context of $e^+e^-$ collisions~\cite{Basham:1978bw,Basham:1978zq}, is an event-shape observable that encodes correlations among multiple particles through their energy-weighted angular separations. 
EECs have become a powerful tool for probing the dynamics of Quantum Chromodynamics (QCD) in a wide range of regimes~\cite{Moult:2025nhu}, including heavy ions~\cite{Chen:2020vvp,Chen:2020adz,Komiske:2022enw,Andres:2023xwr,Andres:2024xvk, Yang:2023dwc,Xing:2024yrb,Apolinario:2025vtx,Liu:2025ufp,Ke:2025ibt,Budhraja:2025ulx,Singh:2024vwb,Barata:2024wsu,Barata:2024ukm,Barata:2025zku,Barata:2023zqg,Barata:2023bhh,Zhao:2025ogc}, transverse momentum dependent (TMD) physics in the back-to-back limit~\cite{Ebert:2020sfi, Kang:2023big, Kang:2024dja}, hadronization process~\cite{Chen:2024nfl, Zhang:2026tjo,Barata:2025uxp}, dihadron fragmentation~\cite{Lee:2025okn,Guo:2025zwb,Chang:2025kgq,Kang:2025zto,Herrmann:2025fqy}, as well as conformal collider physics~\cite{Hofman:2008ar, Belitsky:2013ofa, Belitsky:2013xxa, Korchemsky:2019nzm, Dixon:2019uzg, Lee:2022ige}.
Extensions and variants of EEC have also been shown to provide sensitive probes of the internal dynamics of nuclear matter. In particular, the transverse energy-energy correlator (TEEC) focuses on transverse momenta of produced particles at hadron colliders~\cite{Gao:2019ojf, Ali:2012rn, Li:2020bub, Kang:2023oqj, Kang:2024otf, Kang:2025vjk, Ganguli:2025aqa, Bhattacharya:2025bqa}, while the nucleon energy-energy correlator (NEEC) accesses the internal structure of nucleons at deep inelastic scattering (DIS)~\cite{Liu:2022wop, Liu:2023aqb,Chen:2024bpj,Mantysaari:2025mht,Huang:2025ljp,Cao:2023oef,Cao:2023qat,Li:2023gkh}. 
These works demonstrate that focusing on the energy flow instead of individual hadrons can offer new insights into a wide range of different scattering processes.

As a natural extension of the EEC framework, one may also consider correlating the energy flow with respect to a fixed reference direction, leading to the one-point energy correlator (OPEC)~\cite{Basham:1977iq}. While structurally simpler than the two-point EEC, the OPEC retains sensitivity to the underlying dynamics and offers a complementary perspective on energy flow in scattering processes. 
Recently, OPECs have been used to study a wide range of scattering processes~\cite{Li:2021txc, Kang:2023big, Mi:2025abd, Gao:2025evv, Song:2025bdj, Gao:2025cwy,Zhu:2025qkx, Cao:2025icu, Fu:2025hpc}.

A central goal of the future Electron-Ion Collider (EIC)~\cite{Accardi:2012qut,Aschenauer:2017jsk,Anderle:2021wcy} is to explore the regime of high parton densities in nuclei at ultrarelativistic energies. 
As the collision energy increases, the gluon density grows rapidly due to enhanced gluon radiation, an effect that is amplified in heavy nuclei due to the large number of nucleons. 
Such growth should eventually be tamed down by gluon recombination, leading to an effect known as gluon saturation. 
The high-energy limit is conventionally characterized by a small Bjorken variable $x_B \simeq Q^2/(W^2 + Q^2)\ll 1$, where $Q^2$ is the virtuality of the photon and $W^2$ the center-of-mass energy squared for the photon-nucleon system.
A convenient framework for describing nucleons in this high-energy limit is
the Color Glass Condensate (CGC) effective field theory~\cite{Gelis:2010nm,Weigert:2005us} that describes the gluonic composition of hadrons and nuclei in terms of a classical color field. Within the CGC framework, the evolution of the color field with respect to energy is determined by the Jalilian-Marian--Iancu--McLerran--Weigert--Leonidov--Kovner (JIMWLK) equation~\cite{Iancu:2000hn,Iancu:2001ad,Iancu:2001md,Ferreiro:2001qy,Jalilian-Marian:1996mkd,Jalilian-Marian:1997jhx,Jalilian-Marian:1997qno}, although a numerically simpler form of it, the Balitsky--Kovchegov equation~\cite{Balitsky:1995ub,Kovchegov:1999yj}, is sufficient for many practical applications. 

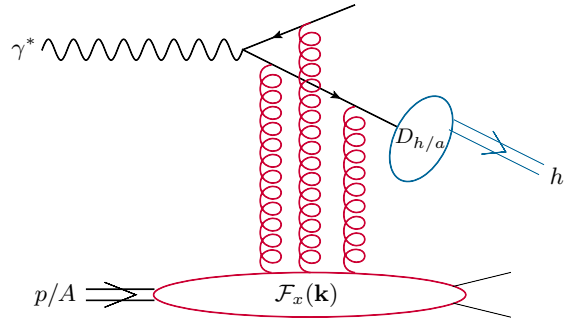
\begin{figure}
\centering
\begin{overpic}[width=0.9\columnwidth]{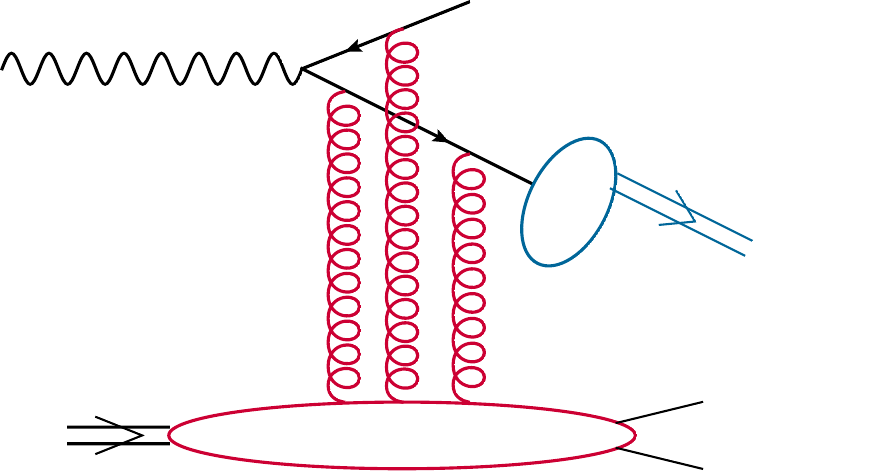}
    \put(40,3){$\mathcal F_x(\kt)$}
    \put(-5,45){$\gamma^*$}
    \put(-1,3){$p/A$}
    \put(87,23){$h$}
    \put(60.5,30){\footnotesize $D_{h/a}$}
\end{overpic}
\caption{
Schematic illustration of the $\gamma^*+p/A\to h+X$ process in deep inelastic $e+p$ (or $e+A$) collisions in the small-$x$ regime.
}
\label{fig:EEC}
\end{figure}

In the small-$x$ regime, energy correlators are infrared-safe observables that are directly sensitive to the underlying dipole amplitude in protons and nuclei~\cite{Liu:2023aqb,Kang:2023oqj,Kang:2025vjk,Bhattacharya:2025bqa,Ganguli:2025aqa,Mantysaari:2025mht}. In electron--proton DIS, the energy flow observable (equivalent to the OPEC) was first introduced in Refs.~\cite{Meng:1991da,Nadolsky:1999kb}. More recently, Ref.~\cite{Li:2021txc} analyzed the back-to-back limit within TMD factorization, establishing its connection to transverse-momentum-dependent parton distributions.

In this work, we study the OPEC in DIS for both proton and gold nuclei under the EIC kinematics in the small-$x$ region. Within the CGC framework, we compute the OPEC in the intermediate-angle regime. As illustrated in Fig.~\ref{fig:EEC}, at leading order the incoming photon splits into a quark--antiquark pair, which subsequently scatters off the gluon shock wave---representing multiple gluon exchanges---of the nuclear target. The outgoing parton  then fragments into an observed hadron. 
 The correlation between the produced hadron and the incoming proton probes the structure of the nuclear target in the small-$x$ regime.
We also compute the nuclear modification factors of the resulting OPECs.

This paper is organized as follows. In \cref{sec:theory}, we derive the expression for the DIS OPEC at small $x$ in the CGC framework. In \cref{sec:results}, we present the numerical results and discuss the associated physical mechanisms. We summarize our results in \cref{sec:conclusion}.

\begin{figure}
\includegraphics[width=0.45\textwidth]{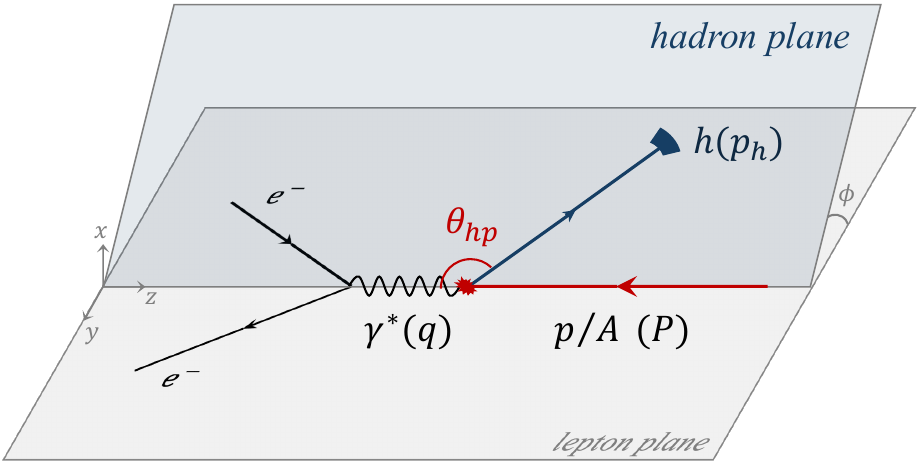} 
\caption{Illustration of EEC for SIDIS in the Breit frame. \label{fig:setup}}
\end{figure}

\section{Theoretical formalism}
\label{sec:theory}
\subsection{DIS one-point energy correlator}
Energy-correlator observables are defined as an energy-weighted sum over all produced hadrons.
For DIS OPEC, this corresponds to measuring the angular distribution of outgoing hadrons with respect to the incoming nucleus, as defined in Ref.~\cite{Li:2021txc},\footnote{In Ref.~\cite{Li:2021txc}, this observable is referred to as DIS EEC. Our definition differs in that it is not normalized to the inclusive cross section. 
}
\begin{equation}
\label{eq:BEEC}
    \frac{\dd{\Sigma}_\lambda}{\dd{\cos\theta}} = \sum_h \int\dd\sigma^{\gamma^*_\lambda+p \to h+X}\, z_h\, \delta(\cos\theta_{hp} - \cos\theta)\;,
\end{equation}
where $\lambda=T, L$ is the photon polarization, and $z_h$ is the longitudinal momentum fraction of the outgoing hadron specified by
\begin{equation}
\label{eq:zh}
    z_h\equiv \frac{P\cdot p_h}{P\cdot q},
\end{equation}
with $P$, $p_h$, and $q$ being the 4-momenta of the incoming proton/nucleus, outgoing hadron $h$, and intermediate virtual photon, respectively. We denote by $\theta_{hp}$ the polar angle between the incoming proton/nucleus and the outgoing hadron.
To simplify the notation, we parameterize the angle into
\begin{equation}
\label{eq:tau}
    \tau = \frac{1 + \cos \theta}{2} \;.
\end{equation}
In this parametrization, the back-to-back limit ($\theta\to\pi$) corresponds to $\tau\to0$, while the collinear limit ($\theta\to0$) corresponds to $\tau\to1$.

We work in the Breit frame in which the virtual photon is moving fast along the positive $z$-direction, while the nucleon moves along the opposite direction, as illustrated in \cref{fig:setup}.
The photon carries momentum 
\begin{equation}
\label{eq:qmu}
    q^\mu=\frac Q{\sqrt2}(\bar n^\mu-n^\mu)\;,
\end{equation}
in which $Q=\sqrt{-q^2}$ is its virtuality.
The light-cone null vectors are defined as $n^\mu\equiv (0,1,0,0)$ and $\bar n^\mu\equiv (1,0,0,0)$.
Throughout the manuscript, we use the light-cone variables expressed as $v^\mu=(v^+, v^-,\boldsymbol{v})=((v^0+v^3)/\sqrt{2}, (v^0-v^3)/\sqrt{2}, v^1, v^2)$. Note that $ v\vdot  n=v^+$ and $ v\vdot \bar n=v^-$, and the boldface notation $\boldsymbol{v}$ denotes transverse components. We also write $v_\perp =|\boldsymbol{v}|$ for the magnitude of the transverse vector.
The incoming proton/nucleus has a 4-momentum given by
\begin{equation}
\label{eq:Pmu}
    P^\mu=\frac{Q}{\sqrt2x_B}n^\mu\;,
\end{equation}
where $x_B\equiv Q^2/(2q\cdot P)$ is the Bjorken variable.

In the dipole picture that is convenient for calculations at small $x$, the incoming virtual photon fluctuates into a quark--antiquark pair that scatters off the target eikonally. 
We keep track of the parton (denoted as $a$) that will further fragment into the produced hadron, while integrating out the other parton. 
We define the longitudinal momentum fraction of the parton $a$ with respect to the virtual photon as
\begin{align}
    \hat \xi\equiv \frac{p_a\cdot n}{q\cdot n}=\frac{p_a^+}{q^+}\;.
\end{align}
In the collinear factorization approach for fragmentation, the produced hadron, $h$, carries a fraction $z$ of the parton's momentum, $p_h=z p_a$.
We assume that the hadron momentum is sufficiently large such that its mass can be neglected.
By imposing the on-shell condition $p_h^+ =p_{h,\perp}^2/(2 p_h^-)$, it is readily seen that the hadron transverse momentum can be determined from $\theta_{hp}$:
\begin{equation}
\label{eq:ph}
    \abs{\pt_h}=z\hat\xi Q\cot\frac{\theta_{hp}}2.
\end{equation}

The cross section for $\gamma^*+p \to h+X$ can be written in terms of the partonic cross sections for $\gamma^*+p \to a+X$ as
\begin{multline}
\label{eq:fragmentation}
    \frac{\dd{\sigma^{\gamma^*+p \to h+X}}}{\dd{z_h}\dd^2\pt_h}=\int_{z_h}^1\frac{\dd{z}}{z^3}\sum_a D_{h/a}(z)
    \frac{\dd{\sigma^{\gamma^*+p \to a+X}}}{\dd{\hat\xi}\dd^2{\pt_a}}\;,
\end{multline}
where $ D_{h/a}(z)$ is the parton-to-hadron fragmentation function, and we note that $z_h=p_h^+/q^+=z\hat\xi$.
The differential partonic cross section is as follows \cite{Marquet:2009ca}:
\begin{align}
\begin{split}
    \label{eq:parton-cross-section}
        \frac{\dd{\sigma^{\gamma^*_\lambda+p \to a+X}}}{\dd{\hat\xi}\dd^2{\pt_a}}=&\frac{2\aem e_f^2N_cS_\perp}{(2\pi)^2}\int\dd^2{\qt}\,\\
    &
       \times  \mathcal H_\lambda( \qt,\pt_a,  \hat\xi,Q^2)
        \mathcal F_x(\qt)\;,
\end{split}
\end{align}
in which for different photon polarizations
\begin{subequations}
    \begin{multline}
     \mathcal H_L( \pt_a, \qt, \hat\xi,Q^2)  \equiv 
4\hat\xi^2(1-\hat\xi)^2
        Q^2\\
        \times\left(\frac1{\pt_a^2+\ef^2}-\frac1{(\pt_a-\qt)^2+\ef^2}\right)^2\;,
    \end{multline}
        \begin{multline}
 \mathcal H_T( \pt_a, \qt, \hat\xi,Q^2) \equiv    [\hat\xi^2+(1-\hat\xi)^2]\\
        \times\abs{\frac{\pt_a}{\pt_a^2+\ef^2}-\frac{\pt_a-\qt}{(\pt_a-\qt)^2+\ef^2}}^2\;.
    \end{multline} 
\end{subequations}
with $\ef^2\equiv \hat\xi(1-\hat\xi)Q^2$. 
Here, $S_\perp$ is the transverse size of the target, the summation over $a$ runs over quarks and antiquarks, and $e_f$ is the dimensionless fractional charge of parton $a$ with flavor $f$ ($\bar f$). The function $\mathcal F_x(\qt)$ denotes the dipole amplitude in the momentum space, related to the S-matrix for dipole–target scattering by
\begin{equation}
    \mathcal{F}_x(\qt) =  \int \frac{ \dd[2]{\rt}}{(2\pi)^2} 
    e^{-i \qt \vdot \rt}
    S_x(\rt),
\end{equation}
where $\rt$ is the dipole transverse separation. We employ the CGC effective field theory to model the dipole amplitude, with its dependence on the $x$-variable governed by the running-coupling Balitsky--Kovchegov (rcBK) equation~\cite{Balitsky:1995ub,Kovchegov:1999yj} with the Balitsky prescription~\cite{Balitsky:2006wa}. We use the same set of parameters as in our previous study of the transverse energy--energy correlator~\cite{Kang:2025vjk}.

For the initial condition of the rcBK equation we use the model~\cite{Casuga:2023dcf}
\begin{multline}\label{eq:MVe}
    S_{x_0}(r) =\\  \exp\qty[-\frac{\qty(r^2  Q^2_{s0})^{\gamma}}{4}\log(\frac{1}{\Lambda_{\rm QCD}  r} + e \cdot e_c )]\;.
\end{multline}
For the proton target, we take the median values of the relevant parameters from the 4-parameter fit (with $\gamma=1$ fixed) in Ref.~\cite{Casuga:2023dcf}. 
For nuclear targets, we modify the saturation scale and the transverse size by~\cite{Kang:2023oqj}
\begin{align}\label{eq:Qs_A}
    Q_{s0,A}^2 = c A^{1/3} Q_{s0}^2\;,
    \qquad
    S_{\perp, A} = A^{2/3} S_\perp / c\;.
\end{align}
The parameter $c$ takes into account the uncertainty in the nuclear geometry, and we vary it between $0.5 < c < 1.0$~\cite{Dusling:2009ni,Tong:2022zwp}.

By inserting the above expressions for the cross section into the OPEC defined in \cref{eq:BEEC}, we obtain
\begin{align}
\begin{split}
    \label{eq:BEEC-substituted}
    \frac{\dd{\Sigma}_\lambda}{\dd{\tau}} = &\sum_{h,a} \int\dd{\hat\xi}\dd^2\pt_h\dd{z}\frac1z D_{h/a}(z)
    \hat\xi\frac{\dd{\sigma^{\gamma^*_\lambda+p \to a+X}}}{\dd{\hat\xi}\dd^2{\pt_a}}\\
    &\times \delta\left(\frac{\abs{\pt_h}^2}{(z\hat\xi Q)^2+\abs{\pt_h}^2} -\tau\right)\;.
\end{split}
\end{align}
Note that after the change of variables $z_h\to \hat\xi$, the lower limit of the $z$ integral becomes $0$.
We apply the delta function in carrying out the $p_{h,\perp}$ integral, thereby fixing the value of $p_{a,\perp}(=p_{h,\perp}/z)$ as well,
\begin{equation}
    \label{eq:BEEC-substituted2}
    \begin{split}
    \frac{\dd{\Sigma}_\lambda}{\dd{\tau}} = &\sum_{h,a} \int_0^1\dd{\hat\xi}\int_0^{2\pi}\dd{\phi}
    \int_0^1\dd{z}
    \frac{z\hat\xi^3Q^2}{2(1-\tau)^2} D_{h/a}(z)\\
    &\times \frac{\dd{\sigma^{\gamma^*+p \to a+X}}}{\dd{\hat\xi}\dd^2{\pt_a}}
    \bigg|_{p_{a,\perp}=\hat\xi Q\sqrt{\tau/(1-\tau)}}
    \;.
    \end{split}
\end{equation}
Here, $\phi=\arg \pt_h=\arg \pt_a$ is the azimuthal angle of the hadron and of the parent parton.
The dependence on the fragmentation functions canceled via the momentum-sum rule
\begin{equation}
\label{eq:momentum-sum}
  \sum_h  \int_0^1 \dd{z} z D_{h/a}(z) = 1\;.
\end{equation}

It follows that,
\begin{equation}
    \label{eq:BEEC-substituted3}
    \begin{split}
    \frac{\dd{\Sigma}_\lambda}{\dd{\tau}} & =
    \frac{2\aem N_cS_\perp Q^2}{(2\pi)^2 (1-\tau)^2}
    \sum_{f}e_f^2
    \int\dd^2{\qt}
       \int_0^{2\pi}
   \dd{\phi}
   \int_0^1
   \dd{\hat\xi}
      \\
    &
    \times   \hat\xi^3
       \mathcal H_\lambda( \qt,\pt_a,  \hat\xi,Q^2)
        \mathcal F_x(\qt)
    \bigg|_{p_{a,\perp}=\hat\xi Q\sqrt{\tau/(1-\tau)}}
    \;,
    \end{split}
\end{equation}
where we account for the antiquark contribution by introducing a factor of $2$, i.e., $\sum_a\to 2\sum_f$, and we include only the light quark flavors $f=u,d,s$ in the sum.
After performing the integrations over the remaining azimuthal angles, the resulting OPEC can be written as,
\begin{multline}\label{eq:EEC_fin}
    \frac{\dd{\Sigma}_\lambda}{\dd{\tau}} = \frac{\aem N_cS_\perp Q^2}{\pi(1-\tau)^2}\sum_{f} e_f^2  \int_0^1\dd{\hat\xi}
    \int_0^\infty\dd{q_\perp}\,q_\perp\,\hat\xi^3\\
    \times\mathcal F_x(q_\perp) h_\lambda\left(q_\perp,\hat\xi Q\sqrt{\frac{\tau}{1-\tau}}, \hat\xi,Q^2\right)\;,
\end{multline}
where the functions $h_\lambda$ for longitudinal and transverse polarizations are given by
\begin{subequations}
\label{eq:hLT}
    \begin{multline}
  h_L(q_\perp,k_\perp, \hat\xi,Q^2)\equiv
\hat\xi^2(1-\hat\xi)^2 Q^2 8\pi\\
\times\left[
 \frac{ a}{(a^2-b^2)^{3/2}}
 -\frac{2 }{(a^2-b^2)^{1/2}d}
 +\frac{1}{d^2}
 \right],  
    \end{multline}
    \begin{multline}
  h_T(q_\perp,k_\perp, \hat\xi,Q^2)\equiv
[\hat\xi^2+(1-\hat\xi)^2] 2\pi\\
\times\left[
  -\frac{\ef^2 a}{(a^2-b^2)^{3/2}}
  +\frac{2\ef^2+a-d}{d(a^2-b^2)^{1/2}}
  -\frac{\ef^2}{d^2}
  \right]\;,
\end{multline}
\end{subequations}
with the shorthand definitions,
\begin{align}
    a\equiv \ef^2 + q_\perp^2 + k_{\perp}^2\,,
    \quad
    b\equiv 2q_\perp k_{\perp}\,,
    \quad
    d \equiv \ef^2 + k_{\perp}^2\,.
\end{align}

In analogy to the inclusive DIS cross section~\cite{Golec-Biernat:1998zce}, the OPEC can be parameterized in terms of different photon polarizations.
We define the structure-function-like OPECs as
\begin{align}\label{eq:EEC2}
    \frac{\dd{\tilde\Sigma_2}}{\dd{\tau}  }
   = \frac{Q^2}{4\pi^2 \aem } 
    \left(
    \frac{\dd{\Sigma_L}}{\dd{\tau}}
    + 
    \frac{\dd{\Sigma_T}}{\dd{\tau}}
    \right)\;,
\end{align}
and 
\begin{align}\label{eq:EEClambda}
    \frac{\dd{\tilde\Sigma_\lambda}}{\dd{\tau}  }
   = \frac{Q^2}{4\pi^2 \aem } 
    \frac{\dd{\Sigma_\lambda}}{\dd{\tau}}
\;.
\end{align}
This rescaling of the OPEC has the advantage of absorbing most of the $Q^2$-dependence and making the quantities dimensionless.
The reduced OPEC, a counterpart of the reduced cross section, can be written as
\begin{align}\label{eq:EECr}
    \frac{\diff \tilde\Sigma_r}{\diff \tau}=
     \frac{\dd{\tilde\Sigma_2}}{\dd{\tau}  }
     -
     \frac{y^2}{\qty[1+(1-y)^2]}
    \frac{\dd{\tilde\Sigma_L}}{\dd{\tau}}
 \;.
\end{align}
To characterize nuclear effects relative to the proton baseline, we define the nuclear modification factor for the OPEC as
\begin{equation}\label{eq:RpA_def}
      R_{A}(\tau) = 
  \left(\dv{\tilde \Sigma_{\lambda,A}}{\tau}\right)
  \bigg/
    \left(A\dv{\tilde \Sigma_{\lambda,p}}{\tau}\right)\;.
\end{equation}
Without nuclear effects, this ratio is normalized to one.

\section{Numerical results}
\label{sec:results}

\begin{figure}[t!]
  \centering 
    \includegraphics[width=0.48\textwidth]{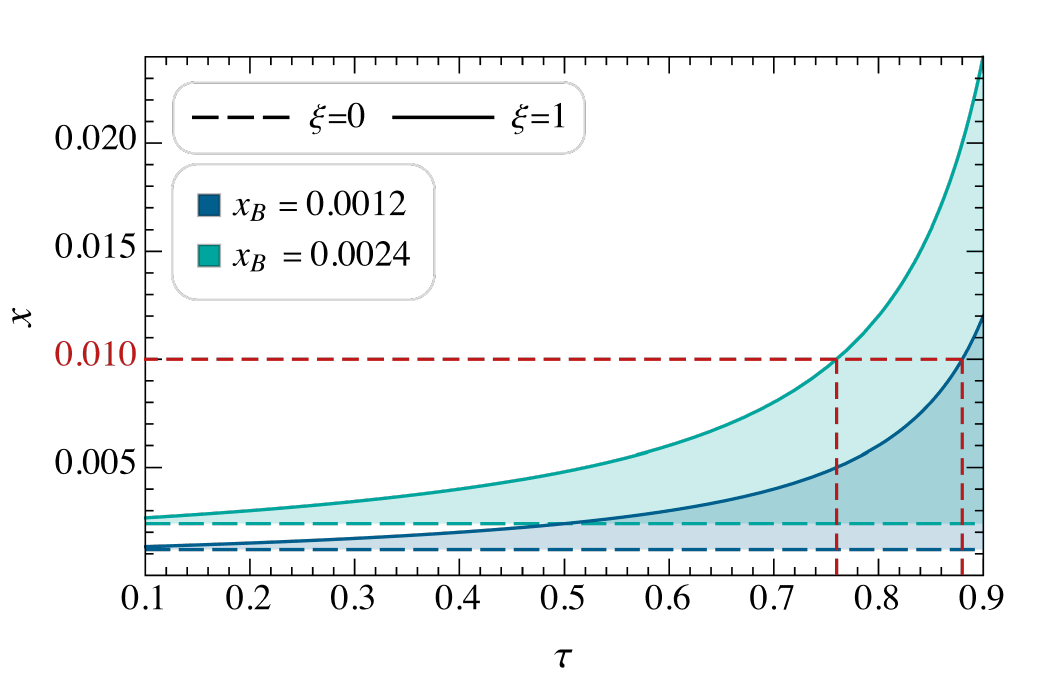} 
  \caption{Illustration of the correspondence between $\tau$ and $x$ at two different values of $x_B$. The dashed (solid) curves correspond to evaluations at $\xi=0$ ($\xi=1$), while the shaded regions indicate the results between these limits. 
\label{fig:tau_xA}  }
\end{figure}

We calculate the OPECs under the EIC kinematic conditions for proton or Au-197 nucleus targets. 
The center-of-mass energy of the $e+p/A$ system is taken as $\sqrt{s}=90 \GeV$, corresponding to the highest center-of-mass energy expected for electron--nucleus collisions at the EIC~\cite{AbdulKhalek:2021gbh}, and the photon virtuality is chosen as $Q=1\sim 5 \GeV$ to probe saturation effects. We consider representative kinematic configurations with the inelasticity $y = Q^2/(x_B s)$ set to $0.5$ and $0.9$, corresponding to moderate and large energy transfer to the hadronic system.
The center-of-mass energy of the $\gamma^*+p/A$ system can be determined as
\begin{align}
    \label{eq:Wsq}
    W^2 = Q^2 \left(\frac{1}{x_B}-1\right)
    = sy - Q^2
    \;,
\end{align}
such that varying the inelasticity at fixed $s$ and $Q^2$ corresponds to varying the energy $W^2$.

The scale $x$ entering the dipole amplitude function in the differential cross section, 
\cref{eq:parton-cross-section}, can be determined by momentum conservation and reads
\begin{equation}
    \label{eq:xscale}
    x=\frac{p_a\cdot \bar n - q\cdot \bar n}{P\cdot \bar n}=\frac{Q^2+\pt_a^2/\hat\xi}{W^2+Q^2}\;.
\end{equation}
Here we have neglected the nucleon mass, which is suppressed by the energy scale.
We have seen that, in the OPEC expression \cref{eq:BEEC-substituted3}, the transverse momentum of the fragmenting parton is given by $p_{a,\perp}=\hat\xi Q\sqrt{\tau/(1-\tau)}$.
Therefore, at a fixed $x_B$, the OPEC at a given $\tau$ is probing the nuclear target at a range of $x$ for $\hat\xi\in[0,1]$,
\begin{align}
    x 
    =x_B\frac{1-(1-\hat\xi)\tau}{1-\tau}\;.
\end{align}
We illustrate this fact at two different values of $x_B$ in \cref{fig:tau_xA}.

\addfignewquadruple{L}{lambda}{T}{lambda}{2}{2}{r}{1}{0.5}
\addfignewdouble{L}{lambda}{r}{r}{3}{0.5}
\addfignewdouble{L}{lambda}{r}{r}{1}{0.9}
\addfignewdouble{L}{lambda}{r}{r}{3}{0.9}

The small-$x$ formalism is applicable only for $x \lesssim  0.01$; accordingly, in practice the dipole amplitude is evaluated in this range and extrapolated beyond it.
Consequently, at a given value of $x_B$, only the range of $\tau$ corresponding to this domain is reliable. The upper bound of this region $\tau_H=1-x_B/0.01$ is indicated by a vertical dashed line in the OPEC plots when applicable.

We present in \cref{fig:Q1_y0.5} the OPECs at $y=0.5$, $Q=\SI{1}{GeV}$, decomposed according to photon polarization into $ \tilde\Sigma_L$, $ \tilde\Sigma_T$, $ \tilde\Sigma_2$, and $ \tilde\Sigma_r$, respectively. 
The corresponding $x_B$ is very small, $2.47\times10^{-4}$.
The longitudinally polarized OPEC, $ \tilde\Sigma_L$, shown in \cref{subfig:L-1-0.5}, exhibits a concave-up behavior in the interval $0.1 \le \tau \le 0.9$ for both proton and gold targets.
The corresponding nuclear modification factor $R_{A}$ is approximately flat between $0.25$ and $0.5$ for $\tau < 0.7$, and then increases to the range $0.5-0.75$ as $\tau$ approaches $0.9$. 
This upward trend indicates a weakening of nuclear suppression toward larger $x$, as expected: the CGC framework predicts that gluon saturation effects are more pronounced at smaller $x$, leading to stronger nuclear modification in that region.
For the gold target, the uncertainty associated with the nuclear geometry (parameterized by $c$) is sizable but remains relatively uniform across the $\tau$ interval.

For the transversely polarized case, $ \tilde\Sigma_T$, shown in \cref{subfig:T-1-0.5}, the proton curve exhibits a milder rise at large $\tau$, whereas the gold results increase steadily over the examined region.
Consequently, $R_{A}$ grows almost monotonically, from roughly $0.2$–$0.35$ at $\tau=0.1$ to about $0.65$–$0.8$ at $\tau=0.9$.
In addition, the overall magnitude of $\tilde\Sigma_T$ is about four to five times larger than that of $\tilde\Sigma_L$. As a result, the combined observables $\tilde\Sigma_2$ and $\tilde\Sigma_r$ are dominated by the transverse contribution and therefore exhibit similar qualitative behavior, as seen in \cref{subfig:2-1-0.5,subfig:r-1-0.5}.
The same pattern persists for $y=0.5, 0.9$ and $Q=1,3~\GeV$, so in the following we present only $\tilde\Sigma_L$ and $\tilde\Sigma_r$ for brevity.

In \cref{fig:Q3_y0.5}, we raise the virtuality to $Q=\SI{3}{GeV}$ while keeping $y=0.5$, which corresponds to a larger Bjorken-$x$, $x_B=2.22\times10^{-3}$. 
In this kinematic regime, the OPEC predominantly probes comparatively larger values of $x$, as indicated by the vertical dashed line marking $x=0.01$.
Both the longitudinal and reduced OPECs decrease with increasing $\tau$. The longitudinal component continues to fall even at large $\tau$ (beyond the $x=0.01$ threshold), whereas the reduced one tends to level off.
In both cases, the proton OPEC decreases more rapidly with $\tau$ than the corresponding gold results, resulting in a steadily increasing nuclear modification factor $R_{A}$. By $\tau=0.9$, $R_{A}$ approaches unity, signaling the disappearance of significant nuclear modification when small-$x$ effects become negligible.
Compared to the results obtained at $Q=1 \GeV$ in \cref{fig:Q1_y0.5}, the OPEC magnitude at small $\tau$ increases, as expected from the larger photon virtuality enhancing the cross section. At the same time, the effect of nuclear modification gets smaller, reflecting that this kinematics probes a region further from the saturation regime as described by the rcBK equation.

We next consider the results at $y=0.9$, shown for $Q=1\GeV$ and $Q=3\GeV$ in \cref{fig:Q1_y0.9} and \cref{fig:Q3_y0.9}, respectively.
The OPECs at $y=0.9$ resemble those at $y=0.5$ with slight quantitative differences.
The longitudinal OPEC values are slightly higher than the corresponding $y=0.5$ cases, with the difference more pronounced at $Q=\SI{3}{GeV}$. In contrast, for the reduced polarizations, the OPEC values are slightly lower than their $y=0.5$ counterparts.

In all cases, the nuclear modification factor $R_{A}<1$, reflecting suppression from small-$x$ effects.
The $\tau$-dependence of the OPEC shows stronger suppression at smaller $\tau$ (smaller momentum scale $p_{a,\perp} \propto Q\sqrt{\tau/(1-\tau)}$ and lower $x$), which weakens as $\tau$ increases.
This is consistent with gluon saturation effects that are more relevant for smaller momenta and lower values of $x$. 
Increasing $Q$ reduces the suppression, highlighting the sensitivity of OPEC to the underlying gluon density.

\section{Conclusion}\label{sec:conclusion}
In this work, within the framework of the Color Glass Condensate effective field theory, we have derived the one-point energy correlator (OPEC) between the produced hadron and the incoming nucleus in high-energy deep inelastic scattering in the small-$x$ regime. Extending previous analyses that focused on the back-to-back limit, we obtained expressions valid at intermediate angles, enabling a systematic study of the angular dependence of energy flow in high-energy DIS. Because the dependence on fragmentation functions cancels via the momentum-sum rule, the only nonperturbative input is the dipole amplitude describing the scattering of a quark--antiquark dipole from the classical color field. Therefore, we propose that the OPEC is a sensitive and promising probe of the gluon distribution at small $x$ which can provide insights on the underlying gluon saturation in nucleons and nuclei.

We presented numerical results for representative kinematic regions relevant to the future Electron--Ion Collider. We find sizable nuclear suppression at small $\tau$, which gradually weakens as $\tau$ increases or as the photon virtuality $Q$ becomes larger. The angular variable $\tau$ effectively scans different values of the longitudinal momentum fraction $x$, with small $\tau$ probing the lowest-$x$ region where saturation effects are strongest. In particular, at large angles (corresponding to small $\tau$ and lower transverse momentum scales), the OPEC exhibits enhanced sensitivity to nuclear modification effects associated with gluon saturation. 
We note that the predicted nuclear suppression is not uniquely attributable to nonlinear gluon dynamics, as frameworks based on nuclear parton distribution functions and nuclear shadowing with linear DGLAP evolution can also produce 
$R_A<1$ at small $x$~\cite{Frankfurt:2011cs,Kang:2011bp,Perepelitsa:2025qpz}. A dedicated comparison between the CGC predictions using the nonlinear BK evolution and such alternative baselines would be valuable for quantifying the discriminating power of the OPEC and is left for future work.

Taken together, these results indicate that measuring the OPEC in DIS at the future EIC would offer a valuable and fragmentation-independent window into gluon saturation dynamics and the small-$x$ structure of nucleons and nuclei across a wide angular range, complementing other saturation-sensitive measurements toward a consistent phenomenological picture of gluon saturation.

\section*{Acknowledgments}
We thank J.~Barata and X.B. Tong for valuable discussions. Z.K., R.K., and J.P. are supported by the National Science Foundation under grant No.~PHY-2515057. This work is also supported by the U.S. Department of Energy, Office of Science, Office of Nuclear Physics, within the framework of the Saturated Glue (SURGE) Topical Theory Collaboration. M.L. is supported by European Research Council under project ERC-2018-ADG-835105 YoctoLHC; by Maria de Maeztu excellence unit grant CEX2023-001318-M and project PID2023-152762NB-I00 funded by MICIU/AEI/10.13039/501100011033; by ERDF/EU; and by Xunta de Galicia (CIGUS Network of Research Centres). M.L. is also supported by new faculty startup funding by Huazhong University of Science and Technology.

\bibliography{references.bib}

\end{document}